# The Structure Function Ratios $F_2^{\mathrm{Li}}/F_2^{\mathrm{D}}$ and $F_2^{\mathrm{C}}/F_2^{\mathrm{D}}$ at small $x$

THE NEW MUON COLLABORATION (NMC)

*Bielefeld University*[1+], *Freiburg University*[2+], *Max-Planck Institut für Kernphysik, Heidelberg*[3+], *Heidelberg University*[4+], *Mainz University*[5+], *Mons University*[6], *Neuchâtel University*[7], *NIKHEF-K*[8++], *Saclay DAPNIA/SPP*[9**], *University of California, Los Angeles*[10], *University of California, Santa Cruz*[11], *Paul Scherrer Institut*[12], *Torino University and INFN Torino*[13], *Uppsala University*[14], *Soltan Institute for Nuclear Studies, Warsaw*[15*], *Warsaw University*[16*]

M. Arneodo[13a)], A. Arvidson[14], B. Badełek[14,16], M. Ballintijn[8c)], G. Baum[1], J. Beaufays[8b)], I.G. Bird[3,8c)], P. Björkholm[14], M. Botje[12d)], C. Broggini[7e)], W. Brückner[3], A. Brüll[2f)], W.J. Burger[12g)], J. Ciborowski[8,16], R. van Dantzig[8], A. Dyring[14], H. Engelien[2h)], M.I. Ferrero[13], L. Fluri[7], U. Gaul[3], T. Granier[9], M. Grosse-Perdekamp[2i)], D. von Harrach[3j)], M. van der Heijden[8d)], C. Heusch[11], G. Igo[10], Q. Ingram[12], K. Janson-Prytz[14k)], M. de Jong[8], E.M. Kabuß[3j)], R. Kaiser[2], H.J.Kessler[2], T.J. Ketel[8], F. Klein[5], S. Kullander[14], U. Landgraf[2], T. Lindqvist[14], G.K. Mallot[5], C. Mariotti[13l)], G. van Middelkoop[8], A. Milsztajn[9], Y. Mizuno[3m)], A. Most[3], A. Mücklich[3], J. Nassalski[15], D. Nowotny[3n)], J. Oberski[8], A. Paić[7], C. Peroni[13], B. Povh[3,4], R. Rieger[5o)], K. Rith[3p)], K. Röhrich[5q)], E. Rondio[15], L. Ropelewski[16], A. Sandacz[15], D. Sanders[r)], C. Scholz[3n)], R. Seitz[5u)], F. Sever[1,8s)], T.-A. Shibata[4], M. Siebler[1], A. Simon[3t)], A. Staiano[13], M. Szleper[15], Y. Tzamouranis[3r)], M. Virchaux[9], J.L. Vuilleumier[7], T. Walcher[5], C. Whitten[10], R. Windmolders[6], A. Witzmann[2], F. Zetsche[3k)]

**Abstract**

We present the structure function ratios $F_2^{\mathrm{Li}}/F_2^{\mathrm{D}}$ and $F_2^{\mathrm{C}}/F_2^{\mathrm{D}}$ measured in deep inelastic muon-nucleus scattering at a nominal incident muon energy of 200 GeV. The kinematic range $10^{-4} < x < 0.7$ and $0.01 < Q^2 < 70$ GeV$^2$ is covered. For values of $x$ less than 0.002 both ratios indicate saturation of shadowing at values compatible with photoabsorption results.

For footnotes see next page.

* Supported by KBN SPUB Nr 621/E - 78/SPUB/P3/209/94.
** Supported by CEA, Direction des Sciences de la Matière.

a) Now also at Dipartimento di Fisica, Università della Calabria, I-87036 Arcavacata di Rende (Cosenza), Italy.
b) Now at Trasys, Brussels, Belgium.
c) Now at CERN, 1211 Genève 23, Switzerland.
d) Now at NIKHEF-H 1009 DB Amsterdam, The Netherlands.
e) Now at University of Padova, 35131 Padova, Italy.
f) Now at MPI für Kernphysik, 69029 Heidelberg, Germany.
g) Now at Université de Genève, 1211 Genève 4, Switzerland.
h) Now at LHS GmbH, 63303 Dreieich, Germany.
i) Now at University of California, Los Angeles, 90024 Ca, USA.
j) Now at University of Mainz, 55099 Mainz, Germany.
k) Now at DESY, Notkestraße 85, 22603 Hamburg, Germany.
l) Now at INFN-Istituto Superiore di Sanità, 00161 Roma, Italy.
m) Now at Osaka University, 567 Osaka, Japan.
n) Now at SAP AG, Neurottstraße 16, 69190 Walldorf, Germany.
o) Now at Comparex GmbH, Gottlieb-Daimler-Straße, 68165 Mannheim, Germany.
p) Now at University of Erlangen-Nürnburg, 91058 Erlangen, Germany.
q) Now at IKP2-KFA, 52428 Jülich, Germany.
r) Now at University of Houston, 77204 Tx, U.S.A.
s) Now at ESRF, 38043 Grenoble, France.
t) Now at New Mexico State University, Las Cruces Nm, U.S.A.
u) Now at Dresden University, 01062 Dresden, Germany.

Since the discovery by the European Muon Collaboration in 1982 that the structure function $F_2$ per nucleon in iron differs significantly from that of a free nucleon [1], nuclear effects on structure functions have been extensively studied both experimentally and theoretically [2].

The behaviour of the ratio $F_2^A/F_2^{\rm D}$ as a function of the Bjorken scaling variable $x$ has been measured for many nuclei in the range $0.0035 < x < 0.8$. For $x < 0.05$ the ratio is smaller than unity; this depletion is referred to as "shadowing". For $0.05 < x < 0.15$ the ratio is larger than unity by a few percent; this is the so-called "enhancement" or "anti-shadowing" region. There is then a fall of the ratio known as the "EMC effect" up to $x \approx 0.6$, followed by a rise due to Fermi smearing thereafter (fig. 1). All these effects increase with atomic number $A$. In the $Q^2$ range so far explored there is little or no evidence of a $Q^2$ dependence.

Recent experiments have concentrated on the very small $x$ region. In particular the Fermilab E665 experiment found that the ratio $F_2^{\rm Xe}/F_2^{\rm D}$ exhibits saturation of shadowing for $x$ smaller than about 0.001 [3].

In this paper we present results on the ratios of structure functions $F_2^A/F_2^{\rm D}$ for the isoscalar nuclei $^6$Li and $^{12}$C. The ratio $F_2^{\rm C}/F_2^{\rm D}$ was previously measured by the SLAC E139 experiment [4] and by NMC [5, 7]. The ratio $F_2^{\rm Li}/F_2^{\rm D}$ was also measured previously by NMC [6, 7]; in that case however it was obtained by combining the ratios $F_2^{\rm Li}/F_2^{\rm C}$ [6] and $F_2^{\rm C}/F_2^{\rm D}$ [5], measured at different muon energies. The data presented here for both Li/D and C/D extend to much smaller values of $x$ than those available so far.

In deep inelastic charged lepton scattering from an unpolarised target the double differential cross section per nucleon can be written, in the one-photon exchange approximation, as

$$\frac{d^2\sigma_{1\gamma}}{dx dQ^2} = \frac{4\pi\alpha^2}{Q^4}\frac{F_2(x,Q^2)}{x}\left[1 - y - \frac{xyM}{2E} + \left(1 - \frac{2m_\mu^2}{Q^2}\right)\frac{y^2}{2}\left(\frac{1+4M^2x^2/Q^2}{1+R(x,Q^2)}\right)\right] \quad .$$

Here $F_2(x,Q^2)$ is the structure function of the nucleon, $x = Q^2/2M\nu$ is the Bjorken scaling variable and $y = \nu/E$ while $E$, $\nu$ and $-Q^2$ are the incident lepton energy, the energy transfer in the laboratory frame and the four momentum squared of the virtual photon, respectively. The mass $M$ is the proton mass, $m_\mu$ the muon mass and $\alpha$ is the electromagnetic coupling constant. The function $R(x,Q^2)$ is the ratio of the longitudinally to transversely polarised virtual photon absorption cross sections. Assuming $R$ to be independent of the nuclear mass, the structure function ratios $F_2^A/F_2^{\rm D}$ are equal to the corresponding cross section ratios. This assumption is supported by data from SLAC [8] and from NMC [9].

## 2 The experiment

The experiment was performed at the M2 muon beam line of the CERN SPS with the NMC spectrometer [10, 11]. The mean incident muon energy was 197 GeV with a $\pm 4$ GeV spread (rms).

Three independent triggers were used. Muon scattering angles larger than 12 mrad were covered by the standard trigger (T1). The small-angle trigger (T2) was sensitive to muons scattered between 5 and 17 mrad. Both triggers covered the region $x < 0.7$, the T2 events having smaller $Q^2$ and $\nu$ at a given $x$ value. Events at larger $x$ were accepted by T1 only. An additional trigger was installed (T14), sensitive to muon scattering angles as small as 1 mrad, extending the kinematic range down to smaller values of $x$. In order to avoid triggering on divergent beam tracks, only the central part of the beam was used. In this way an extension of the kinematic range down to $x = 10^{-4}$ was achieved, more than an order of magnitude lower than in earlier NMC measurements. Figure 2 shows the kinematic regions covered by the three triggers.

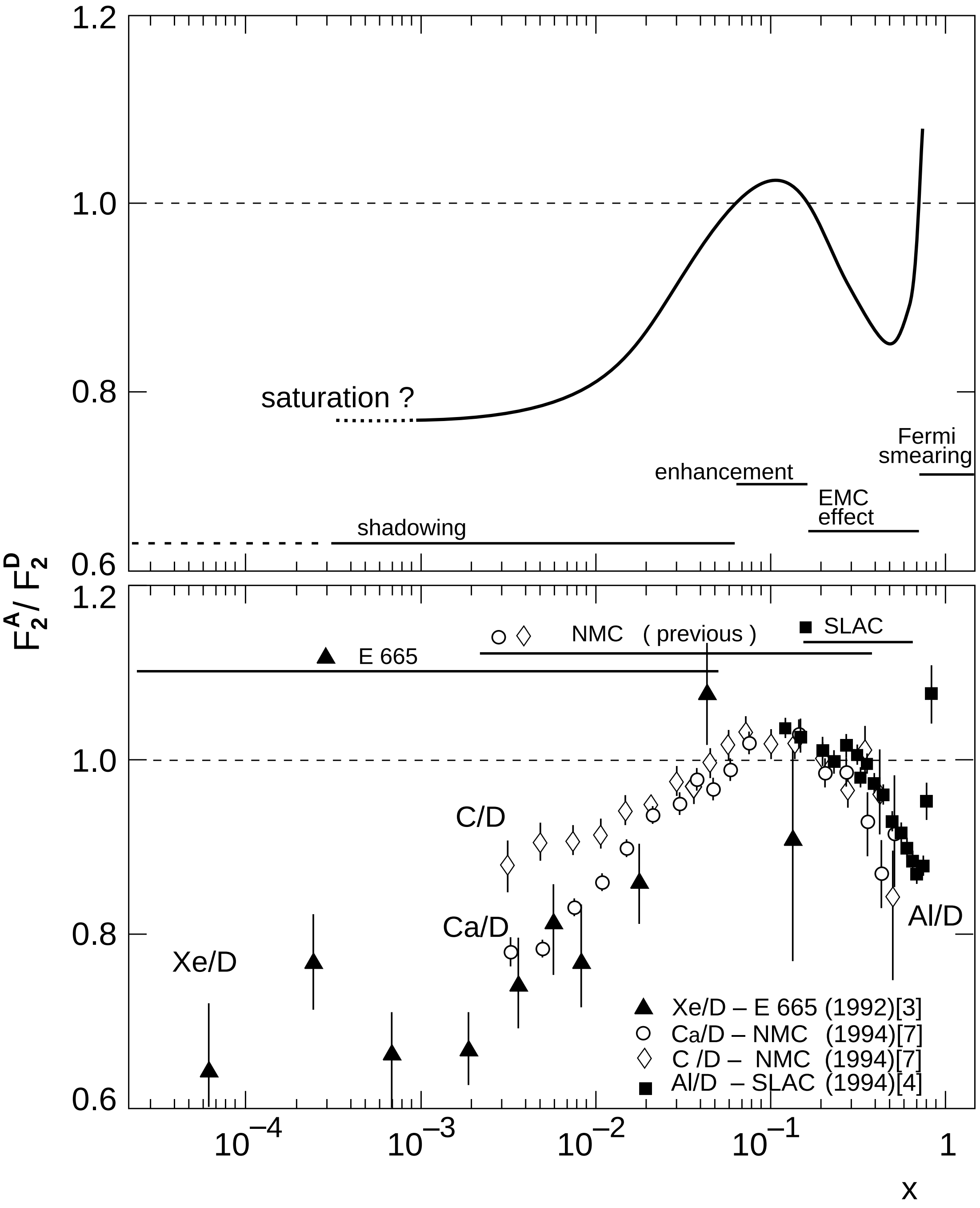

Figure 1: Knowledge about nuclear effects on $F_2$ prior to this work (see text). A phenomenological curve (top) and some recent high precision data sets (below).

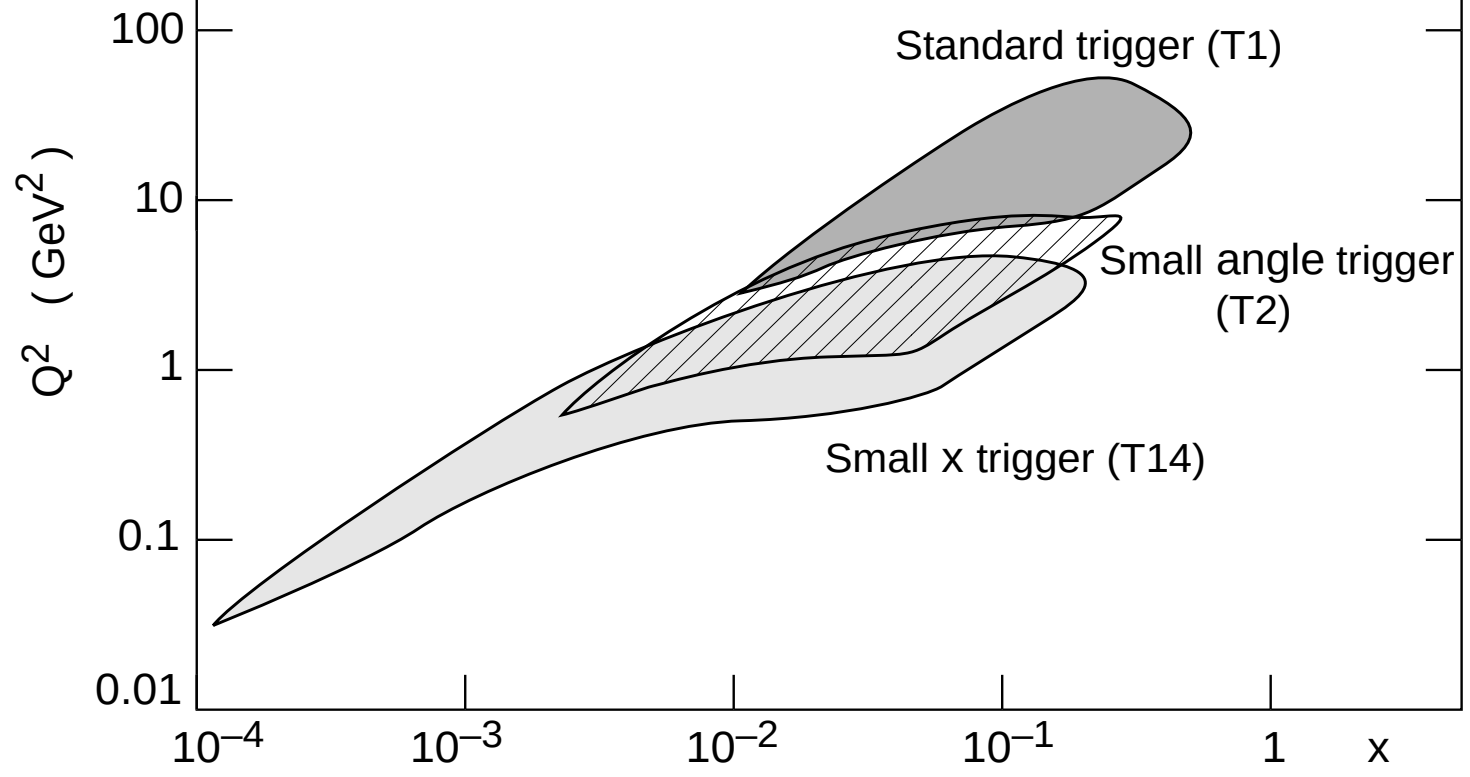

Figure 2: Kinematic regions covered by the physics triggers.

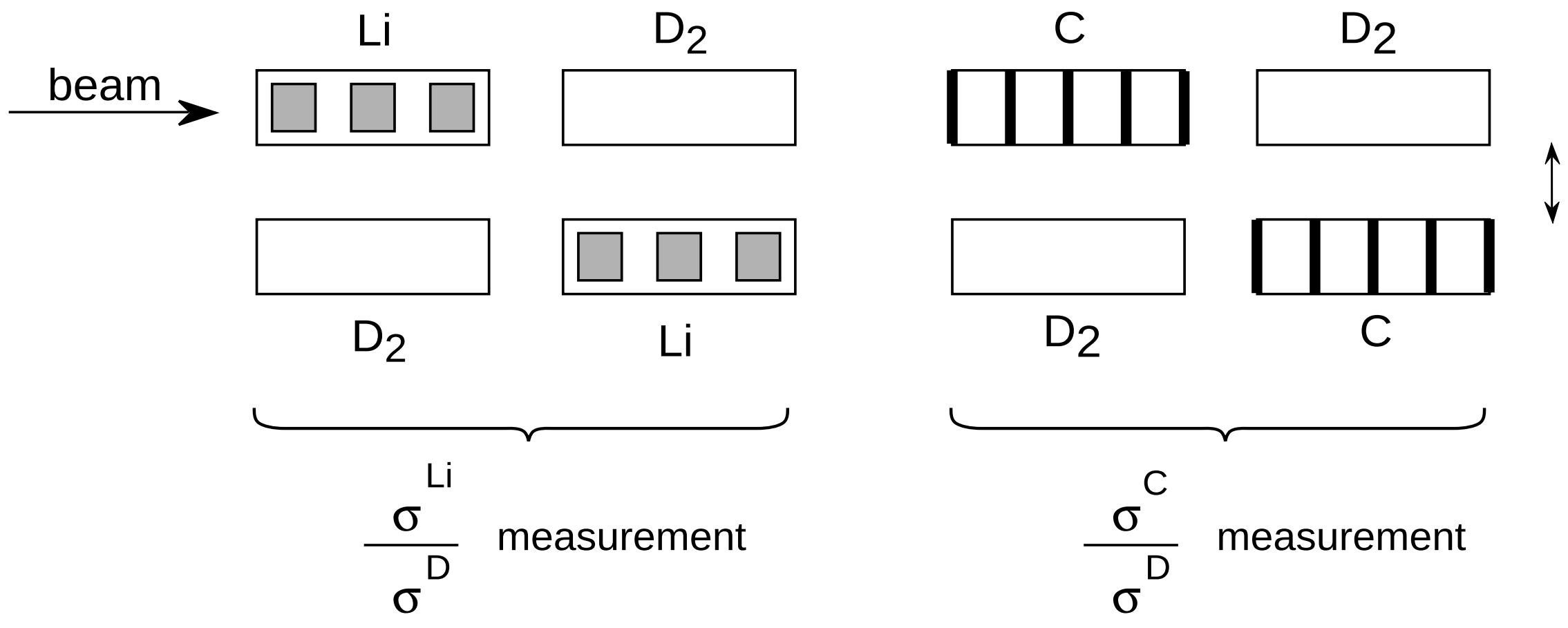

Figure 3: Complementary target setup (see text).

The data discussed in this paper were collected with the target configuration shown in fig. 3. This configuration consisted of two complementary target sets, an upstream one with Li and D targets and a downstream one with C and D targets. Only one of the two target rows was used at a time, simultaneously exposing to the beam one lithium and one deuterium target in the upstream set, and one carbon and one deuterium target in the downstream one. The row was frequently exchanged with the complementary one where the “heavy” target material A (Li or C) and deuterium were interchanged. Geometrical acceptance and efficiency corrections thus cancel in the calculation of the ratios, as do the beam fluxes. The frequent exchange of the two target sets (every 60 min, i.e. about 150 times during the whole run) substantially eliminated the effects of possible time dependences. The cross section ratio then depends only on the number of events $N$ and the number of nucleons $T$ per unit area in the upstream ($u$) and downstream ($d$) targets:

$$\left(\frac{\sigma^A}{\sigma^{\rm D}}\right)_{meas} = \sqrt{\frac{N_u^A \cdot N_d^A}{N_u^{\rm D} \cdot N_d^{\rm D}} \Big/ \frac{T_u^A \cdot T_d^A}{T_u^{\rm D} \cdot T_d^{\rm D}}} \quad . \tag{1}$$

Table 1: Kinematic cuts applied in the standard analysis. A blank space indicates that no cut was applied. $\theta$ is the muon scattering angle.

| variable | | Standard trigger | Small angle trigger | Small $x$ trigger |
|---|---|---|---|---|
| $\theta^{min}$ [mrad] | Li/D | 12 | 5 | 2 |
| $\theta^{max}$ [mrad] | Li/D | | 17 | 12 |
| $\theta^{min}$ [mrad] | C/D | 13 | 6 | 2.65 |
| $\theta^{max}$ [mrad] | C/D | | 19 | 12 |
| $\nu^{min}$ [GeV] | | 20 | 15 | 10 |
| $x^{min}$ | | | | 0.0008 |
| $y^{max}$ | | 0.8 | 0.85 | 0.8 |

With this setup, two ratios were measured simultaneously: $\sigma_{Li}/\sigma_D$ with the upstream targets and $\sigma_C/\sigma_D$ with the downstream ones. The targets had a similar number of interaction lengths for all materials, providing optimal statistical accuracy. In order that the acceptance for the heavy targets be the same as for the deuterium targets, the former were segmented to cover the same length. The liquid deuterium targets were 1 m long; the corresponding lithium ones consisted of three 13 cm long cylinders of $^6$Li kept in an argon atmosphere in plexiglas containers; the cylinders were equally spaced over a 1 m distance. Likewise the carbon targets consisted each of five uniformly spaced 2 cm thick disks in a dallite tube also distributed over a 1 m distance. The total target thicknesses were 17.7 g/cm$^2$, 17.4 g/cm$^2$ and 18.7 g/cm$^2$ for deuterium, lithium and carbon, respectively. The carbon targets were of natural isotopic composition. The deuterium targets contained 1.5 % H atoms from a contamination by HD molecules, while the $^6$Li targets contained 4.5 % $^7$Li.

## 3 Analysis

The data were processed using a chain of programmes [11] which performed pattern recognition and geometrical reconstruction of the incident and scattered muons and of the secondary charged particles.

Cuts were applied to remove events from regions with rapidly changing acceptance and poor vertex resolution (small scattering angle), poor kinematic resolution (small $\nu$) and high background from hadronic decays (low scattered muon energy). The region at high $y$, where radiative corrections are large, was also excluded. At this stage, the region $x < 0.0008$ had to be discarded because it was dominated by $\mu e^-$ elastic scattering events. Table 1 summarizes the cuts applied for the three triggers; we shall refer to them in the following as the "standard cuts". They rely solely on the incoming and scattered muon kinematics.

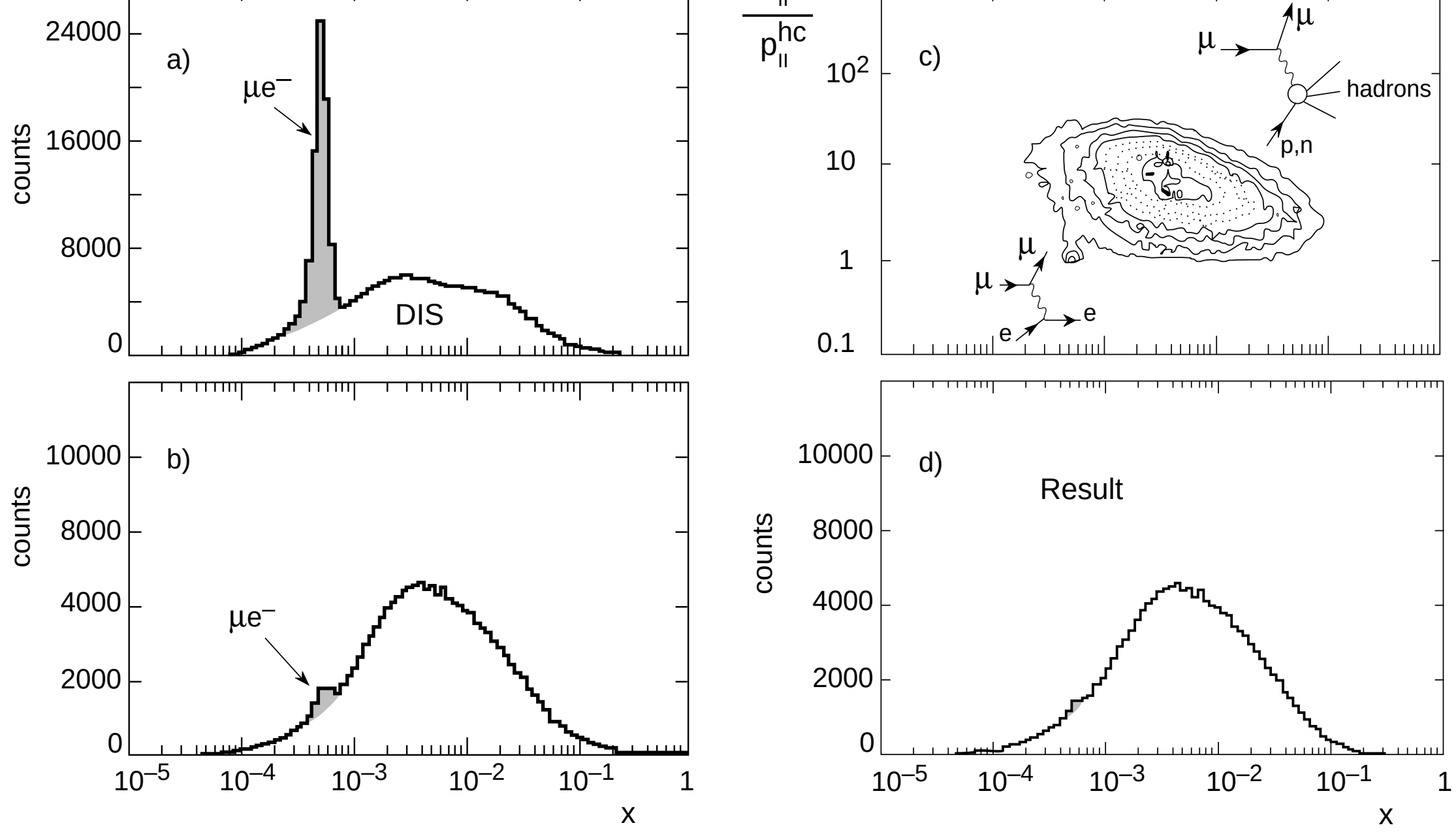

Figure 4: The effects of the hadron cuts. a) The original $x$ distribution dominated by the $\mu e^-$ elastic peak at small $x$. b) The $x$-distribution after requiring a charged track in the spectrometer not identified as an electron or positron by the calorimeter. c) Contour plot of the event distribution over $p_{||}^{\gamma^*}/p_{||}^{hc}$ and $x$ ("hc" stands for hadron candidate). The isolated island around $p_{||}^{\gamma^*}/p_{||}^{hc} = 1$ and $x = m_e/M = 5.4 \cdot 10^{-4}$ represents the two-body $\mu e^-$ scattering events. d) The $x$-distribution with the two-body events removed.

The elastic $\mu e^-$ scattering events occur at $x = m_e/M$, where $m_e$ is the electron mass (fig. 4a). In order to remove this background we used a second set of cuts called the "hadron cuts". At least one hadron with energy larger than 5 GeV fitted to the muon vertex was required. Hadrons, which are the signature of a deep inelastic event, were identified and distinguished from fast electrons (signature of an elastic $\mu e^-$ event) by means of the ratio of energies deposited in the electromagnetic and hadronic parts of the calorimeter H2 [12]. However, due to an acceptance gap in H2 near the beam, about 3% of the electrons were misidentified as hadrons (fig. 4b).

An improvement of the $\mu e^-$ event rejection was achieved by exploiting the kinematics of these events. Elastic $\mu e^-$ scattering is a two-body process, where the electron carries the full momentum lost by the muon (i.e. the momentum of the virtual photon), while in deep inelastic scattering this momentum is shared by several final state particles. We quantify this criterion by observing the ratio of the longitudinal momentum of the virtual photon $p_{||}^{\gamma^*}$, is visible, with a broad continuum of DIS events elsewhere. This made it possible to apply a box cut in the $p_{||}^{\gamma^*}/p_{||}^{hc}$, $x$ plane. The $x$ distribution after this cut is shown in fig. 4d.

The contamination by $\mu e^-$ events in the two affected $x$-bins was estimated to be less than 6% in the final sample, the residue being due to electrons which radiate a real photon. The requirement of a fast hadron in the final state removed all coherent ($\mu-$nucleus) and quasi-elastic ($\mu-$nucleon) events. Radiative corrections, otherwise very large in this low-$x$ region, were thus significantly reduced. Since the additional hadron track fitted to the vertex significantly improves the vertex resolution, the $\theta$ cut could be dropped.

In summary, at least one charged hadron, i.e. a non-muon particle with less than 75% of its energy deposited in the electromagnetic module of the calorimeter was required; for events in the range $0.0003 < x < 0.0008$, the hadron was required to carry less than 2/3 of the virtual photon momentum. To complete the hadron cuts, a higher minimum $\nu$ cut of 60 GeV was imposed for the C/D targets (see section 4.1). The hadron cuts were only applied to T14 data.

The number of events after applying the standard cuts for Li/D is $0.22 \cdot 10^6$, $0.37 \cdot 10^6$ and $0.12 \cdot 10^6$ for T1, T2, and T14 respectively, giving a total of $0.70 \cdot 10^6$. For C/D we have $0.14 \cdot 10^6$, $0.26 \cdot 10^6$ and $0.07 \cdot 10^6$ events in T1, T2 and T14 respectively, totalling $0.47 \cdot 10^6$. After applying the hadron cuts, we are left with 67000 T14 events for Li/D, and 33000 for C/D, but in a different kinematic region going down to $x = 10^{-4}$. Most of the suppression of the T14 events by the hadron cuts is due to the removal of radiative events.

In order to correct the measured yields for coherent and quasielastic scattering and higher order electromagnetic processes, the cross section ratio was calculated from eq. (2) after weighting each event with the radiative correction factor $\eta = \sigma_{1\gamma}/\sigma_{meas}$. These radiative corrections were computed according to the prescription of Akhundov *et al.* [13]. The procedure corrects for the radiative tails of coherent elastic scattering from nuclei and of quasi-elastic scattering from nucleons, as well as for the inelastic radiative tails. As we discussed above, the corrections for the coherent and quasi-elastic tails were not necessary for the sample of data for which a fast hadron was required in the final state.

Several input parameters are needed for the evaluation of the radiative correction factors. The evaluation of the inelastic higher order processes requires the knowledge of both $R$ and $F_2$ over a large range of $x$ and $Q^2$. A fit [14] to the $F_2^{\rm D}$ results of deep inelastic scattering experiments was used in the measured region ($Q^2 > 0.5$ GeV$^2$). For the extrapolation of $F_2$ to lower $Q^2$ various models were used. That by Badełek and Kwieciński [15] combining the generalized vector meson dominance and partonic approaches was used for the radiative correction itself; extrapolations based on the Regge parametrisation by Donnachie and Landshoff [16] and one resulting from a phenomenological fit [11] to low energy data in the resonance region [17], were used to estimate the systematic uncertainty. The structure functions $F_2^{\rm C}$ and $F_2^{\rm Li}$ were obtained by multiplying $F_2^{\rm D}$ by empirical fits to the present cross section ratios. For carbon the SLAC-E139 data [4] for $x > 0.4$ were included in the fit. Since the measured ratio was needed as an input, an iterative procedure was required. The function $R(x, Q^2)$ was assumed to be nucleus independent and taken from a fit [18].

For the quasi-elastic tails, the nucleon form factor parametrisation of Gari and Krümpelmann [19] was used. The reduction of the elastic nucleon cross section with respect to the free nucleon one (quasi-elastic suppression) was evaluated using the results of a calculation by Bernabeu [20] for deuterium and carbon, whereas for lithium the result of Moniz [21]

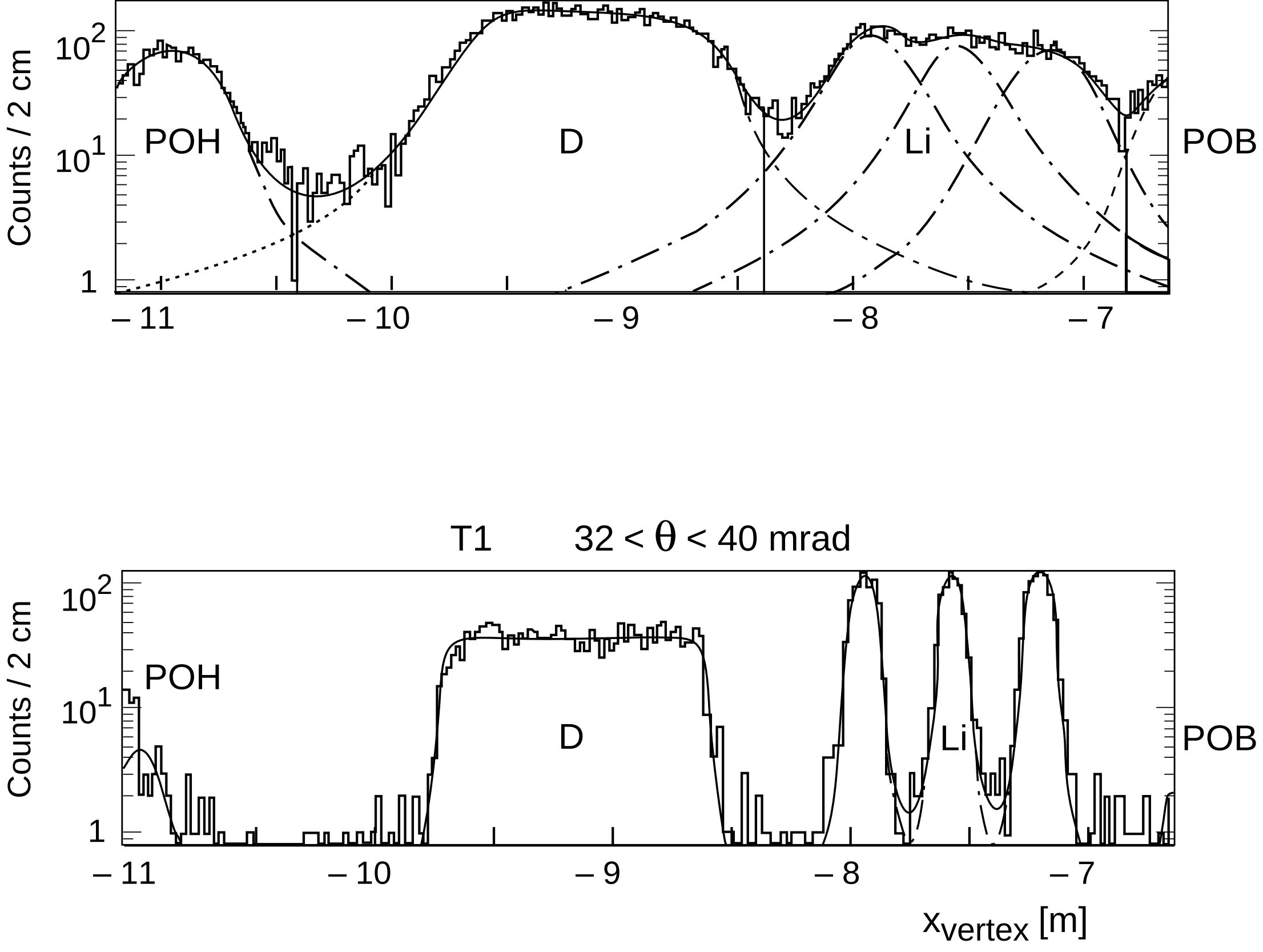

Figure 5: The upstream part of the vertex distribution in two bins – a low $\theta$, low resolution T14 bin (up) and a high resolution T1 bin (down). The events originate (left to right) from a proportional wire chamber P0H, the deuterium target, the three slabs of the segmented lithium target, and from another proportional chamber, P0B. The dashed lines represent the contributions from the individual target segments to the overall fit (full line).

was used.

Finally, in order to evaluate the coherent radiative tails the knowledge of the nuclear elastic form factors is necessary. Parametrisations of available data were used for deuterium [22] and for the heavy targets [23, 24].

For all the above input parameters, alternative parametrisations from the literature were used to estimate the associated systematic uncertainties. For a more detailed discussion, see ref. [10].

With the standard cuts the radiative correction factors $\eta$ for the individual targets range from 0.33 at small $x$ to 1.17 at large $x$ for carbon. However, the resulting mean corrections to the ratios are 15 % and 33 % at the lowest $x$ bin for Li/D and C/D, respectively, and they decrease rapidly with increasing $x$ to less than 2 % for $x > 0.008$. For the hadron cuts the corrections to the ratio were everywhere smaller than 1 %.

The finite resolution of the spectrometer leads to an uncertainty in the position of the interaction vertex. Consequently events might be wrongly attributed to a target material or might fall outside of the target region. To estimate the number of such events, the vertex lowest and highest $\theta$ bin together with the fitted curves. These were used to determine the tails of the vertex distributions. Correction factors which accounted for the wrong target association were calculated from the value of $\theta$ of each event. The largest correction to the ratio from this source was 1.4 %, at the lowest scattering angles.

The target thicknesses were corrected to account for scattering in the air and argon between the lithium and carbon slices, and in the mylar windows of the liquid targets. Holes were found in the lithium slabs, mapped and then convoluted with the beam profile and thus accounted for. A similar procedure was applied to account for the rounded edges of the deuterium target vessels. A correction was also applied to take into account the slight non-isoscalarity of the targets. Finally, the effects of the limited spectrometer resolution in $x$ and $Q^2$ (kinematic smearing effects) on the ratio were estimated by a calculation convoluting the resolution in the scattered muon energy and scattering angle with the multiple Coulomb scattering in the target. The total effect of all these corrections on the ratios was less than 2%.

# 4 Results

## 4.1 $x$ dependence

For both target pairs, the data from the three triggers were analysed separately and the results were found to agree in the region of overlap and therefore combined. Figures 6 and 7 show the results for $F_2^{\rm Li}/F_2^{\rm D}$ and $F_2^{\rm C}/F_2^{\rm D}$; the ratios are shown separately for the standard and for the hadron cuts.

We observe good agreement between the ratio determined with the standard cuts and with the hadron cuts, where the radiative corrections are very small, thus providing a cross check between the two methods. Because of the loss of statistics with the hadron cuts, we use the hadron points only for the lowest four $x$ bins which are inaccessible with the standard cuts.

Figures 8 and 9 show the $x$ dependence of the structure function ratios, averaged over $Q^2$. The systematic errors with the *standard cuts* are dominated at small $x$ by the uncertainties in the radiative corrections. These uncertainties were estimated by varying the input parameters to the radiative corrections program as discussed above. Other contributions to the systematic error include the uncertainty in the vertex smearing and kinematic smearing corrections, the latter being dominant at high $x$. The uncertainty on the measurement of the incoming and outgoing muon momenta ($\delta p/p = 0.2\%$) was also taken into account. An additional systematic error of 0.5% is included over the whole $x$ range as an estimate of the uncertainty of the complementary target method used to extract the ratios. The individual contributions were added in quadrature to obtain the total systematic error.

The systematic errors with the *hadron cuts* have negligible contributions from the radiative corrections. However, the requirement to detect a hadron in the final state might bias the sample due to hadron losses. These may occur due to a reinteraction in the target nucleus or to a collision with a nucleus further downstream.

In both cases a bias to the ratio can result if the effects are different for different target materials. The intranuclear reinteraction is more probable in large nuclei. In ref. [25] it is shown that the differential multiplicity of hadrons created in copper is suppressed by up to

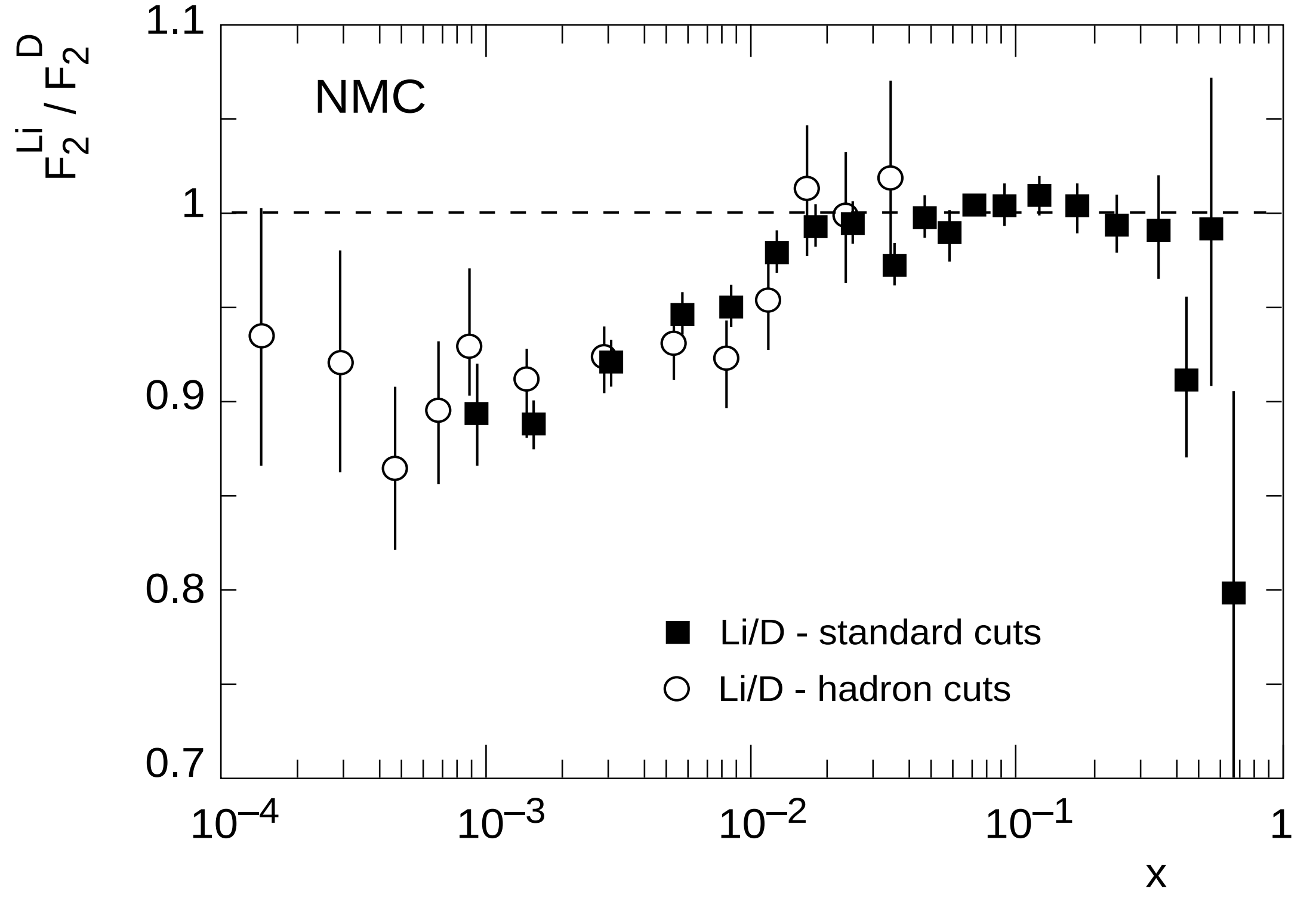

Figure 6: The merged results for $F_2^{\mathrm{Li}}/F_2^{\mathrm{D}}$ from all three triggers (full boxes) compared to the ratios obtained with the hadron cuts on T14 data (open points). The errors shown are statistical only.

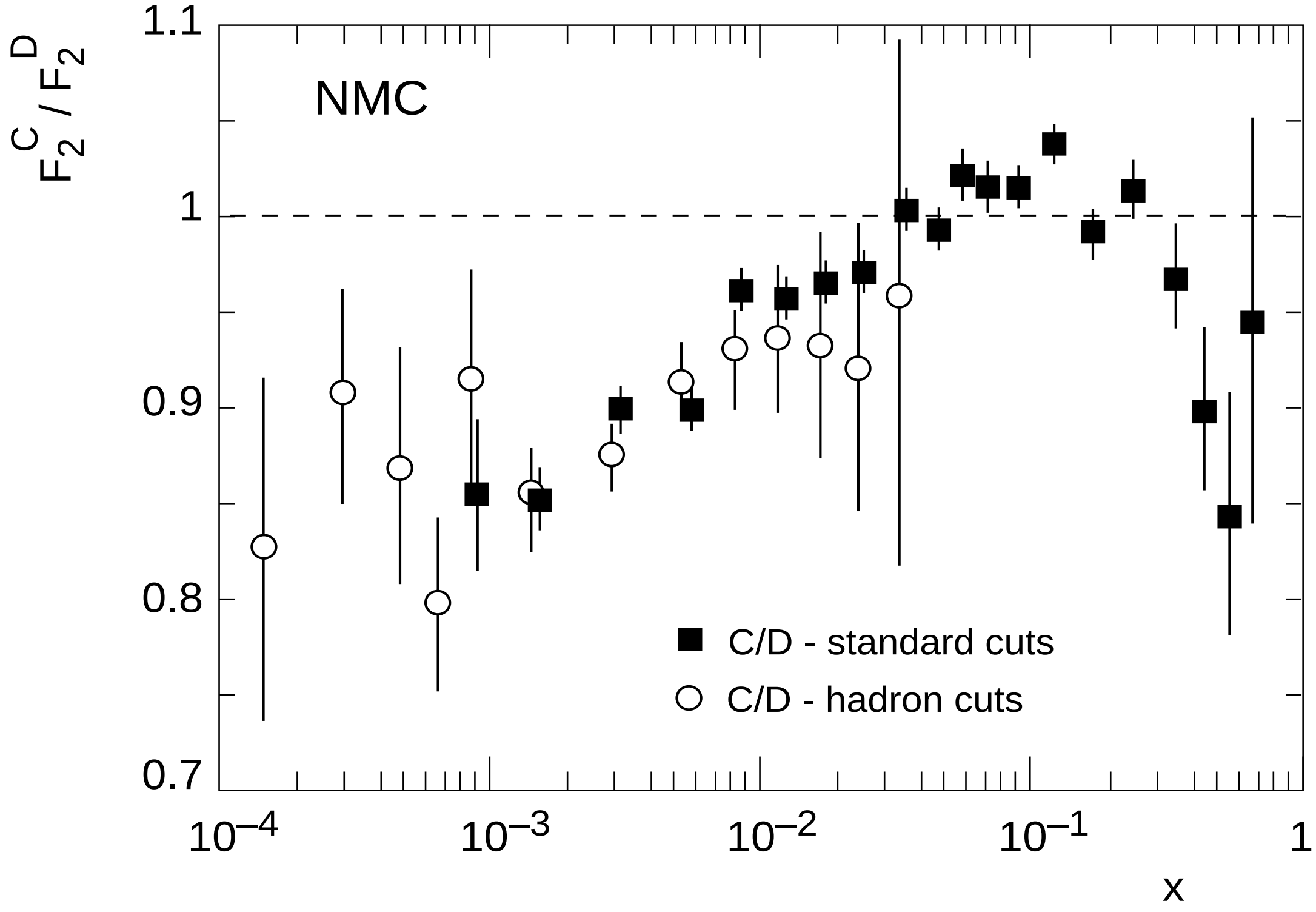

Figure 7: The merged results for $F_2^{\rm C}/F_2^{\rm D}$ from all three triggers (full boxes) compared to the ratios obtained with the hadron cuts on T14 data (open points). The errors shown are statistical only.

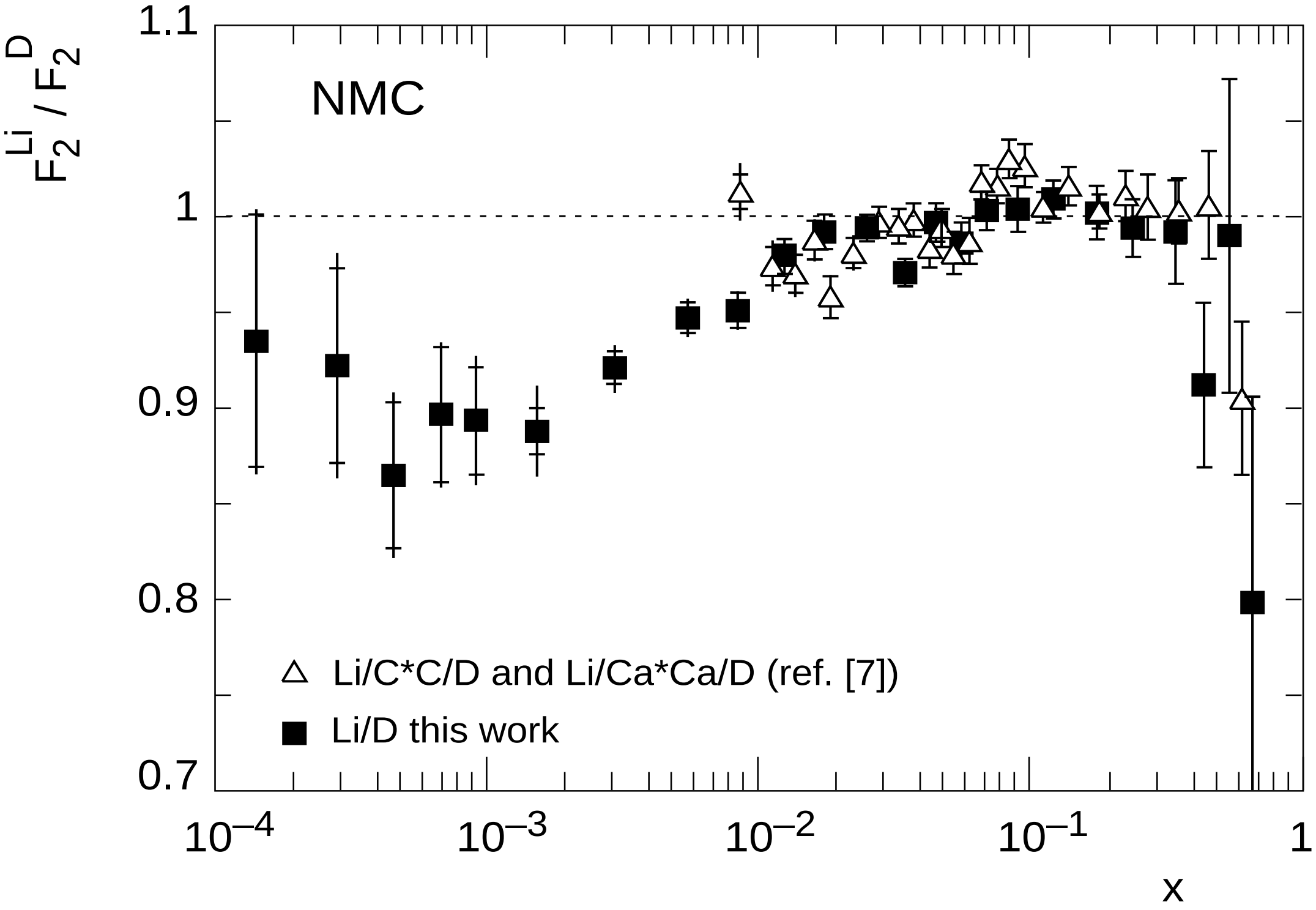

Figure 8: The present result for $F_2^{\rm Li}/F_2^{\rm D}$ compared to the indirect result obtained by multiplying $F_2^{\rm Li}/F_2^{\rm C}$ and $F_2^{\rm Li}/F_2^{\rm Ca}$ (90 GeV) with $F_2^{\rm C}/F_2^{\rm D}$ and $F_2^{\rm Ca}/F_2^{\rm D}$ (200 GeV), respectively [7]. The inner error bars represent the statistical errors, the outer ones the statistical and systematic errors added in quadrature. The relative normalisation uncertainty between the two results, not included in the figure, is 0.8%.

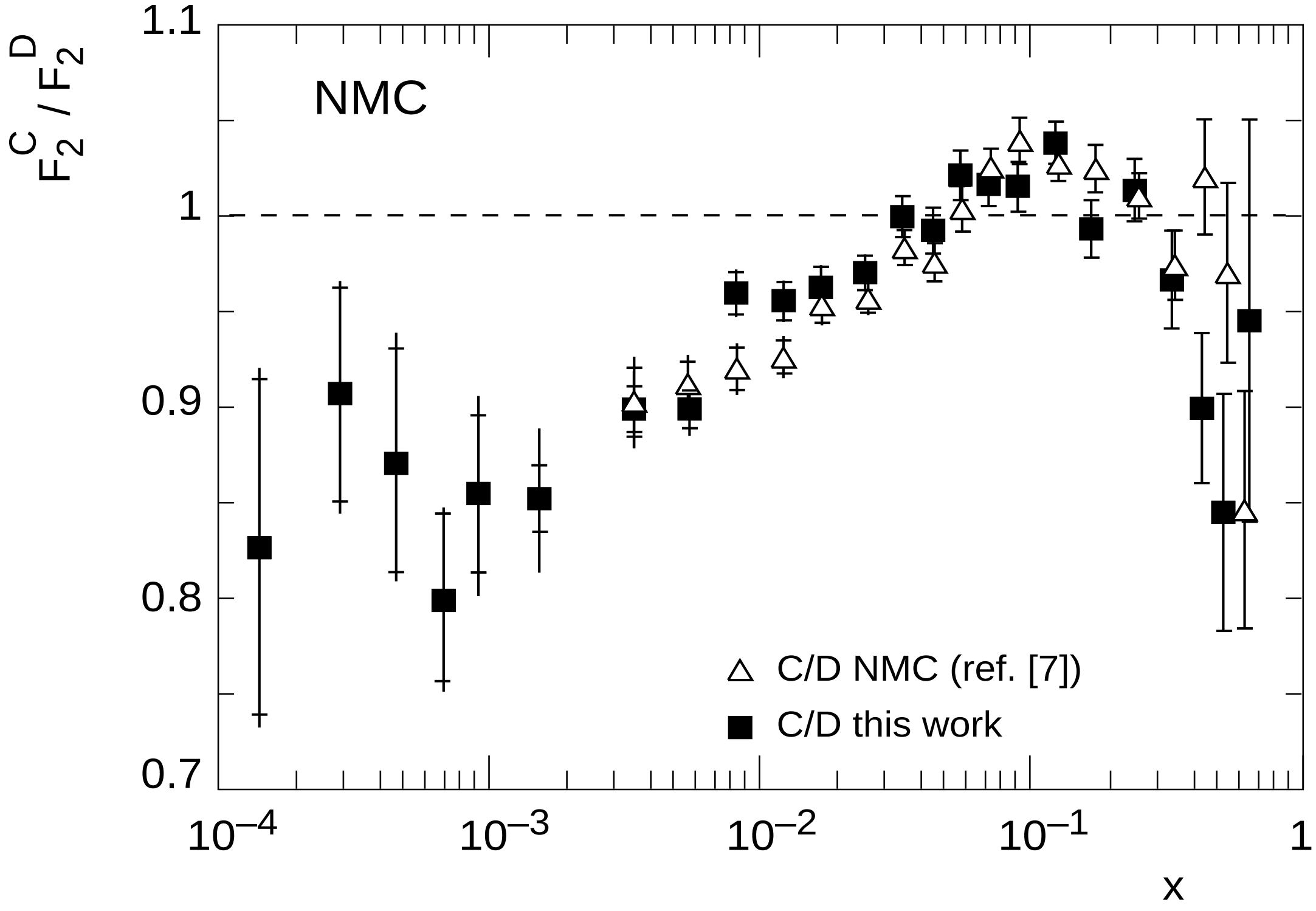

Figure 9: The present result for $F_2^{\rm C}/F_2^{\rm D}$ compared to that of ref. [7]. The inner error bars represent the statistical errors, the outer ones the statistical and systematic errors added in quadrature. The relative normalisation uncertainty between the two results, not included in the figure, is 0.5%.

Table 2: The ratio $F_2^{\rm Li}/F_2^{\rm D}$ and the logarithmic slopes $\partial(F_2^{\rm Li}/F_2^{\rm D})/\partial(\ln Q^2)$ as a function of $x$ from this work. The $x$ value, the $Q^2$ range and mean $Q^2$ (in units of GeV$^2$), and the mean $y$ are also given. The logarithmic slopes for the first five bins are averaged to increase the statistical significance. The normalisation error on the ratios is 0.004 and is not included in the systematic uncertainty.

| $x$ | $Q^2_{min}-Q^2_{max}$ | $< Q^2 >$ | $< y >$ | $\frac{F_2^{\rm Li}}{F_2^{\rm D}}$ ± stat ± syst | $\frac{\partial(F_2^{\rm Li}/F_2^{\rm D})}{\partial(\ln Q^2)}$ ± stat |
|---|---|---|---|---|---|
| $1.4 \cdot 10^{-4}$ | 0.015–0.055 | 0.034 | 0.64 | 0.935±0.066±0.021 | 0.02 ±0.10 (averaged over first five bins) |
| $2.8 \cdot 10^{-4}$ | 0.04–0.09 | 0.066 | 0.65 | 0.922±0.051±0.030 | |
| $4.5 \cdot 10^{-4}$ | 0.04–0.15 | 0.11 | 0.64 | 0.865±0.038±0.022 | |
| $6.7 \cdot 10^{-4}$ | 0.06–0.22 | 0.15 | 0.61 | 0.897±0.036±0.016 | |
| $9.0 \cdot 10^{-4}$ | 0.15–0.25 | 0.21 | 0.67 | 0.894±0.027±0.021 | |
| $1.5 \cdot 10^{-3}$ | 0.15–0.45 | 0.34 | 0.63 | 0.888±0.012±0.022 | −0.01 ±0.05 |
| $3.5 \cdot 10^{-3}$ | 0.15–1.2 | 0.61 | 0.61 | 0.921±0.009±0.010 | −0.005±0.030 |
| $5.5 \cdot 10^{-3}$ | 0.3–2.0 | 1.0 | 0.56 | 0.947±0.008±0.008 | 0.004±0.021 |
| $8.5 \cdot 10^{-3}$ | 0.3–3.0 | 1.4 | 0.51 | 0.951±0.009±0.006 | 0.021±0.020 |
| $1.25 \cdot 10^{-2}$ | 0.3–5.0 | 1.8 | 0.47 | 0.980±0.008±0.006 | 0.016±0.016 |
| $1.75 \cdot 10^{-2}$ | 0.3–6.0 | 2.3 | 0.41 | 0.992±0.009±0.005 | −0.004±0.017 |
| $2.5 \cdot 10^{-2}$ | 0.5–9.0 | 2.8 | 0.36 | 0.994±0.007±0.005 | 0.020±0.012 |
| $3.5 \cdot 10^{-2}$ | 0.7–12 | 3.6 | 0.31 | 0.971±0.008±0.005 | 0.007±0.013 |
| $4.5 \cdot 10^{-2}$ | 0.8–13 | 4.2 | 0.28 | 0.997±0.010±0.005 | 0.014±0.015 |
| $5.5 \cdot 10^{-2}$ | 1.0–16 | 4.8 | 0.27 | 0.986±0.011±0.005 | 0.016±0.017 |
| $7.0 \cdot 10^{-2}$ | 1.2–20 | 6.0 | 0.25 | 1.003±0.010±0.005 | −0.011±0.015 |
| $9.0 \cdot 10^{-2}$ | 1.5–27 | 7.2 | 0.24 | 1.004±0.012±0.005 | 0.006±0.019 |
| 0.125 | 2.0–37 | 9.2 | 0.23 | 1.009±0.010±0.005 | −0.017±0.017 |
| 0.175 | 3.0–50 | 12 | 0.23 | 1.002±0.014±0.005 | −0.036±0.026 |
| 0.25 | 4.0–50 | 17 | 0.22 | 0.992±0.015±0.005 | −0.048±0.034 |
| 0.35 | 10–65 | 23 | 0.21 | 0.990±0.027±0.005 | 0.10 ±0.07 |
| 0.45 | 15–65 | 29 | 0.20 | 0.910±0.043±0.005 | 0.03 ±0.13 |
| 0.55 | 20–90 | 33 | 0.15 | 0.988±0.082±0.006 | 0.13 ±0.36 |
| 0.65 | 25–90 | 39 | 0.13 | 0.801±0.108±0.014 | −0.5 ±0.8 |

Table 3: The merged NMC results for the ratio $F_2^{\rm C}/F_2^{\rm D}$ and the logarithmic slopes $\partial(F_2^{\rm C}/F_2^{\rm D})/\partial(\ln Q^2)$ as a function of $x$. Results from the present work are averaged with the ones from ref. [7]. The $x$ value, the $Q^2$ range and mean $Q^2$ (in units of GeV$^2$), and the mean $y$ are also given. The logarithmic slopes for the first five bins are averaged to increase the statistical significance. The normalisation error on the ratios is 0.004 and is not included in the systematic uncertainty.

| $x$ | $Q^2_{min}-Q^2_{max}$ | $< Q^2 >$ | $< y >$ | $\frac{F_2^{\rm C}}{F_2^{\rm D}}$ ± stat ± syst | $\frac{\partial(F_2^{\rm C}/F_2^{\rm D})}{\partial(\ln Q^2)}$ ± stat |
|---|---|---|---|---|---|
| $1.5 \cdot 10^{-4}$ | 0.015−0.055 | 0.035 | 0.65 | 0.826±0.088±0.033 | −0.24 ±0.15 (averaged over first five bins) |
| $3.0 \cdot 10^{-4}$ | 0.04−0.12 | 0.072 | 0.65 | 0.906±0.057±0.027 | |
| $4.8 \cdot 10^{-4}$ | 0.06−0.15 | 0.11 | 0.64 | 0.870±0.060±0.026 | |
| $6.7 \cdot 10^{-4}$ | 0.08−0.22 | 0.16 | 0.64 | 0.799±0.045±0.025 | |
| $9.0 \cdot 10^{-4}$ | 0.15−0.25 | 0.22 | 0.67 | 0.854±0.041±0.034 | |
| $1.5 \cdot 10^{-3}$ | 0.18−0.5 | 0.36 | 0.63 | 0.852±0.018±0.033 | −0.14 ±0.08 |
| $3.5 \cdot 10^{-3}$ | 0.3−1.2 | 0.70 | 0.61 | 0.899±0.010±0.017 | 0.002±0.039 |
| $5.5 \cdot 10^{-3}$ | 0.3−2.0 | 1.1 | 0.56 | 0.904±0.008±0.010 | 0.004±0.027 |
| $8.5 \cdot 10^{-3}$ | 0.3−3.0 | 1.6 | 0.53 | 0.939±0.008±0.008 | 0.002±0.025 |
| $1.25 \cdot 10^{-2}$ | 0.3−5.0 | 2.2 | 0.50 | 0.939±0.006±0.006 | −0.036±0.016 |
| $1.75 \cdot 10^{-2}$ | 0.3−7.0 | 2.9 | 0.46 | 0.957±0.007±0.006 | 0.013±0.015 |
| $2.5 \cdot 10^{-2}$ | 0.5−10 | 3.6 | 0.40 | 0.963±0.006±0.006 | 0.025±0.011 |
| $3.5 \cdot 10^{-2}$ | 0.7−12 | 4.5 | 0.36 | 0.990±0.007±0.005 | 0.012±0.012 |
| $4.5 \cdot 10^{-2}$ | 0.8−15 | 5.5 | 0.34 | 0.983±0.008±0.005 | 0.014±0.012 |
| $5.5 \cdot 10^{-2}$ | 1.0−20 | 6.4 | 0.32 | 1.011±0.009±0.005 | 0.013±0.013 |
| $7.0 \cdot 10^{-2}$ | 1.0−25 | 7.8 | 0.31 | 1.021±0.007±0.005 | 0.008±0.011 |
| $9.0 \cdot 10^{-2}$ | 1.5−26 | 9.6 | 0.30 | 1.028±0.009±0.005 | 0.009±0.014 |
| 0.125 | 2.0−45 | 12 | 0.28 | 1.032±0.007±0.005 | 0.014±0.012 |
| 0.175 | 3.0−46 | 15 | 0.26 | 1.011±0.010±0.005 | −0.004±0.017 |
| 0.25 | 5.5−65 | 20 | 0.23 | 1.010±0.010±0.005 | −0.002±0.020 |
| 0.35 | 12−90 | 27 | 0.21 | 0.971±0.015±0.005 | −0.040±0.033 |
| 0.45 | 15−90 | 32 | 0.20 | 0.975±0.024±0.005 | −0.07 ±0.05 |
| 0.55 | 20−90 | 37 | 0.17 | 0.925±0.037±0.006 | 0.20 ±0.12 |
| 0.65 | 25−90 | 41 | 0.14 | 0.873±0.053±0.011 | 0.32 ±0.18 |

quark hadronise as it is leaving the nucleus. For this reason, in the hadron method for the C/D ratio we apply a minimum $\nu$ cut of 60 GeV. This brings the methods into agreement. The lowest four $x$ bins, which are the only ones in the final data set determined with the hadron method, are not affected by this cut and no systematic error was assigned for this.

As to the collisions with other nuclei, also no correction was made but a detailed study was done [10]. Calculating the probability of each detected hadron to have been absorbed in flight, we investigated by how much the structure function ratio would have changed if we had weighted events inversely proportionally to the survival probability of the hadron. Combinations with multiple hadrons in the initial and final states were taken into account in three different ways, in particular to account for the hadrons that are missing in the final state. As this calculation was not a long chain Monte-Carlo result, and as it gave small effects to either side of the uncorrected value, depending on the method used, the effects observed were not used to apply a correction, but rather to estimate the systematic uncertainty by taking the maximal effect observed on each bin. This is the dominant contribution to the systematic error in the hadron method.

The normalisation uncertainty on the ratios is not included in the errors shown in figs. 8 and 9. It is due to the uncertainties on the target thicknesses and amounts to 0.4 % and 0.3 % for the $F_2^{\rm Li}/F_2^{\rm D}$ and $F_2^{\rm C}/F_2^{\rm D}$ ratios, respectively.

At small $x$, the results for the structure function ratios are sensitive to the assumption $\Delta R = R^A - R^{\rm D} = 0$. For instance, if $\Delta R = \pm 0.04$, then the ratios change by up to $\pm 0.01$ at small $x$.

The data cover the range $0.0001 < x < 0.7$ and $0.01 < Q^2 < 70$ GeV$^2$. Our results show the well known pattern of shadowing and enhancement. They extend, however, to smaller values of $x$ than previous results, by nearly two orders of magnitude for Li/D and by one for C/D (figs. 8 and 9).

The consistency of the present results with those of ref. [7] was checked using the $\chi^2$ and the run [26] tests and combining [27] the confidence levels of these two independent tests, the first measuring the magnitudes, and the second the signs of the deviations. For the two $F_2^{\rm C}/F_2^{\rm D}$ ratios, the combined confidence level is 3.5%, whereas for the $F_2^{\rm Li}/F_2^{\rm D}$ ratios it is 7.2% if the first bin from ref. [7] at $x = 0.0085$, showing a 4.9 standard deviation discrepancy, is excluded.

The results for $F_2^{\rm Li}/F_2^{\rm D}$ are presented in table 2, and those for $F_2^{\rm C}/F_2^{\rm D}$ combined with the results of ref. [7] are given in table 3.

Figure 10 shows a compilation of small $x$ data on Xe/D, Li/D and C/D, including the ratios at $x = 0$ derived from real photoabsorption experiments [28].

The shadowing in lithium is found to saturate at a value of 0.890±0.010(stat.)±0.021(syst.) for $x < 0.002$, determined by averaging the data below $x = 0.002$. In the same region, the shadowing in carbon saturates at 0.851±0.014(stat.)±0.029(syst.). The present data are consistent with the findings of the E665 collaboration that the Xe/D ratio [3] also saturates for $x < 0.002$, at a value of $0.72 \pm 0.03$(stat.)±0.07(syst.).

These saturation values are in agreement with the results from photoabsorption experiments. At an incident photon energy of 60 GeV it was found that $(\sigma(\gamma C \rightarrow X)/\sigma(\gamma D \rightarrow X)) = 0.832 \pm 0.010$ [28]. Similar data are not available for lithium, but we can estimate the ratio by power-law interpolation (as suggested by [29]) between deuterium and carbon. A weak energy dependence of shadowing has been observed in Cu [28] and the energy of the virtual photon in our data is higher than 60 GeV. The arrows in fig. 10 show how the real photon points would be affected by an analogous energy dependence for C and Li.

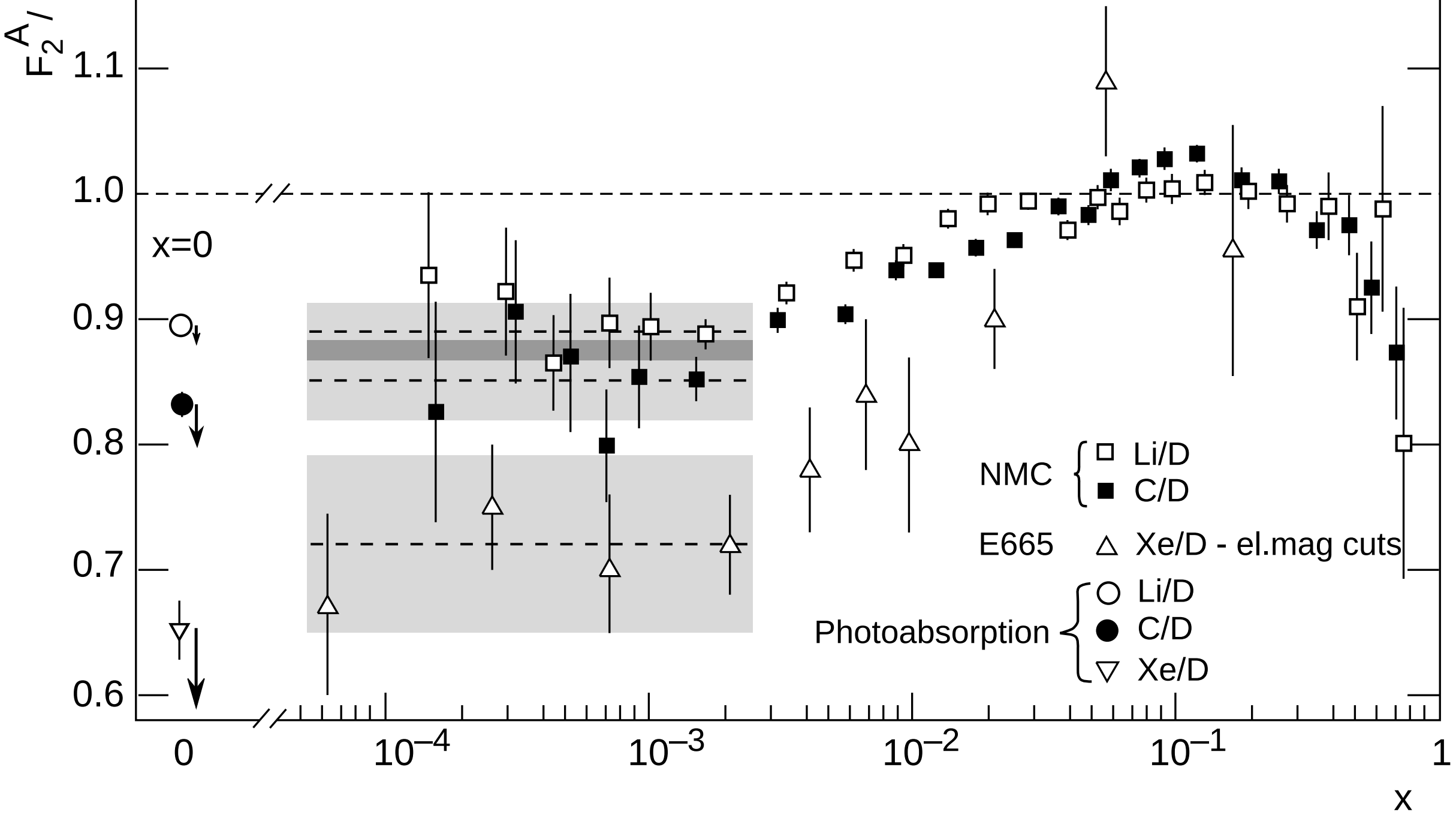

Figure 10: The NMC ratios $F_2^{\rm Li}/F_2^{\rm D}$ (present work) and $F_2^{\rm C}/F_2^{\rm D}$ (merged data, see text), shown with data on $F_2^{\rm Xe}/F_2^{\rm D}$ from the E665 experiment [3]. The points at $x = 0$ show the ratio $\sigma(\gamma A \rightarrow X)/\sigma(\gamma{\rm D} \rightarrow X)$ from photoabsorption experiments at $E_\gamma = 60$ GeV [28]; the arrows indicate by how much these points would move under the energy dependence assumptions described in the text. Dash-dotted lines indicate the saturation values, the shaded areas the errors. The errors on the DIS data points are statistical only; the saturation bands include statistical and systematic errors added in quadrature.

### 4.2 $Q^2$ dependence

In order to study the $Q^2$ dependence of the structure function ratios, the latter were evaluated in bins of $x$ and $Q^2$ (figs. 11-12). For each $x$ bin straight lines in ln $Q^2$ were fitted through the data points in order to extract logarithmic slopes $\partial(F_2^A/F_2^{\rm D})/\partial(\ln Q^2)$. The slopes thus obtained are shown in figs. 13 and 14. The slopes from the lowest five $x$ bins were averaged to increase the significance of the result. The slopes in each $x$ bin are statistically compatible with no $Q^2$ dependence.

## 5 Summary and conclusions

In summary, we have measured nuclear effects on the structure function $F_2$ in lithium and carbon, particularly at small $x$. Compared to earlier experiments, the kinematic range is extended down to $x = 10^{-4}$. The increase of shadowing and antishadowing with the atomic number $A$, as well as the lack of significant $Q^2$ dependence in the measured range is confirmed. In the new kinematic region at very small $x$, we have found an indication of saturation of shadowing at values compatible with the results of photoabsorption experiments.

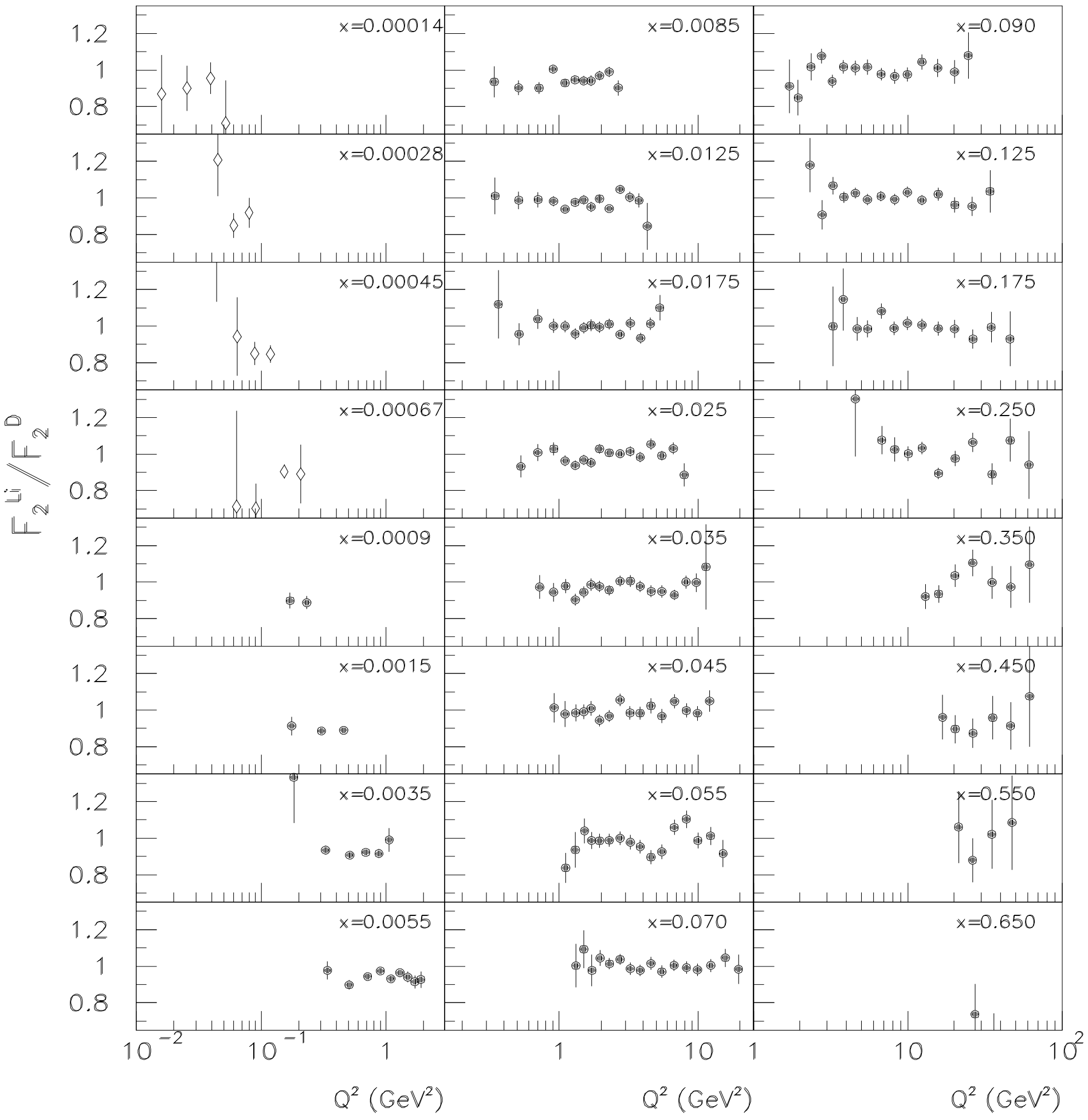

Figure 11: The ratio $F_2^{\mathrm{Li}}/F_2^{\mathrm{D}}$ versus $Q^2$ in $x$ bins. The mean $x$ of each bin is indicated in the figure. The diamonds correspond to T14 data analysed with the hadron cuts, and the black points represent the results obtained by merging the data from all triggers analysed with the standard cuts. The errors shown are statistical.

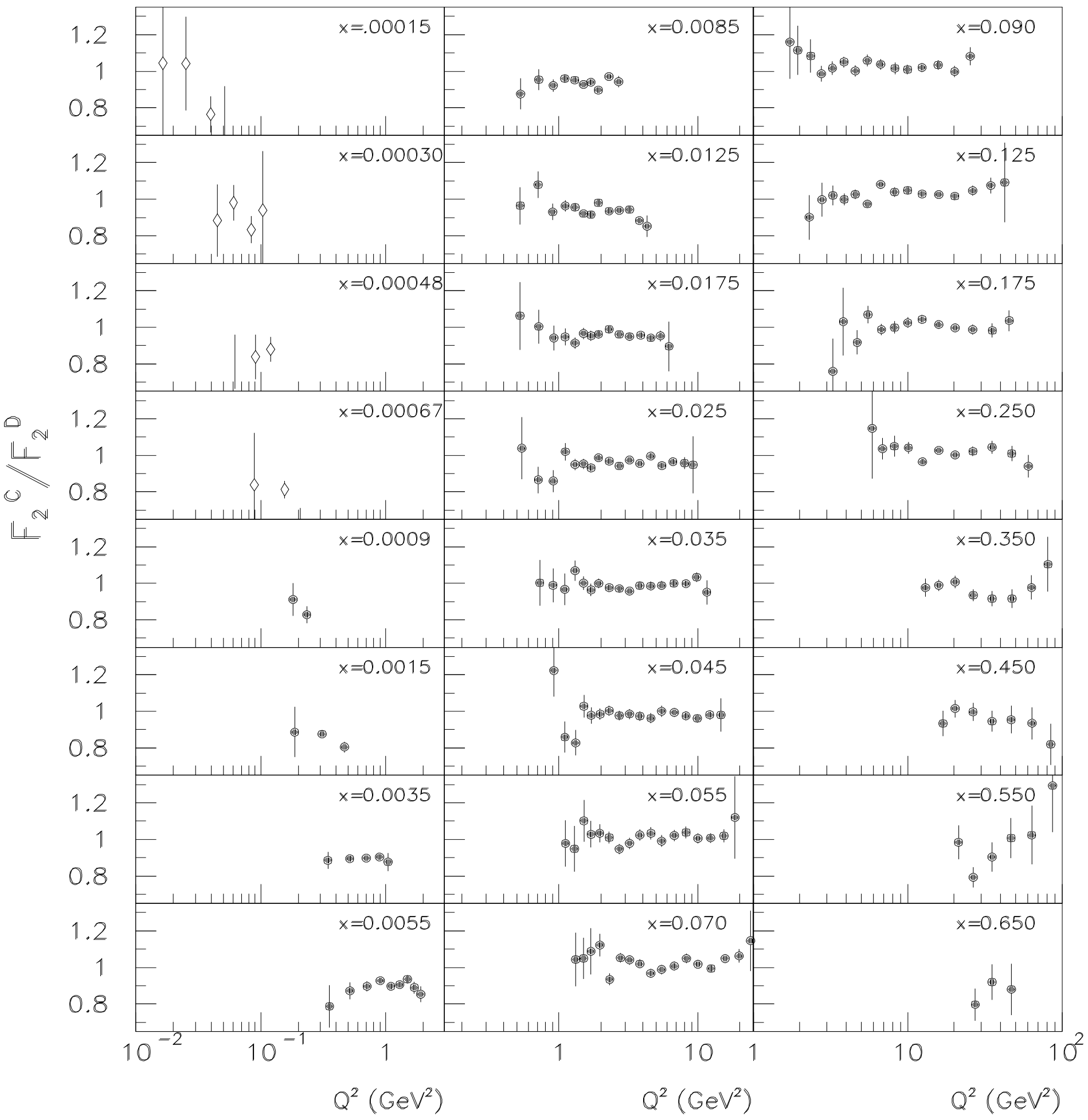

Figure 12: The merged NMC results for the ratio $F_2^{\mathrm{C}}/F_2^{\mathrm{D}}$ versus $Q^2$ in $x$ bins. The mean $x$ of each bin is indicated in the figure. The diamonds correspond to T14 data analysed with the hadron cuts, and the black points represent the results obtained by merging the data from all triggers analysed with the standard cuts together with those of ref. [7]. The errors shown are statistical.

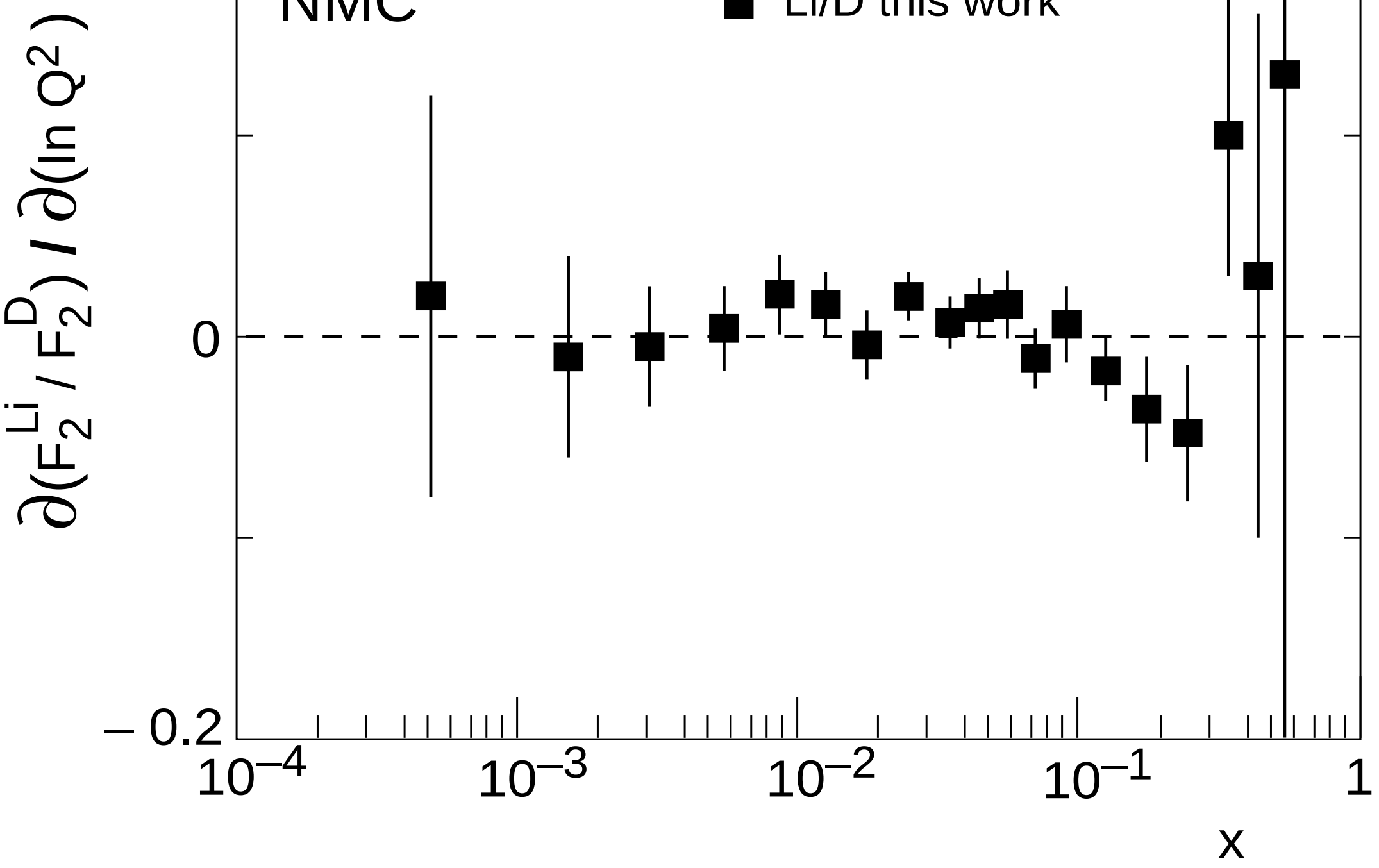

Figure 13: The logarithmic slopes $\partial(F_2^{\rm Li}/F_2^{\rm D})/\partial(\ln Q^2)$ as obtained from linear least squares fits to the points of fig. 11.

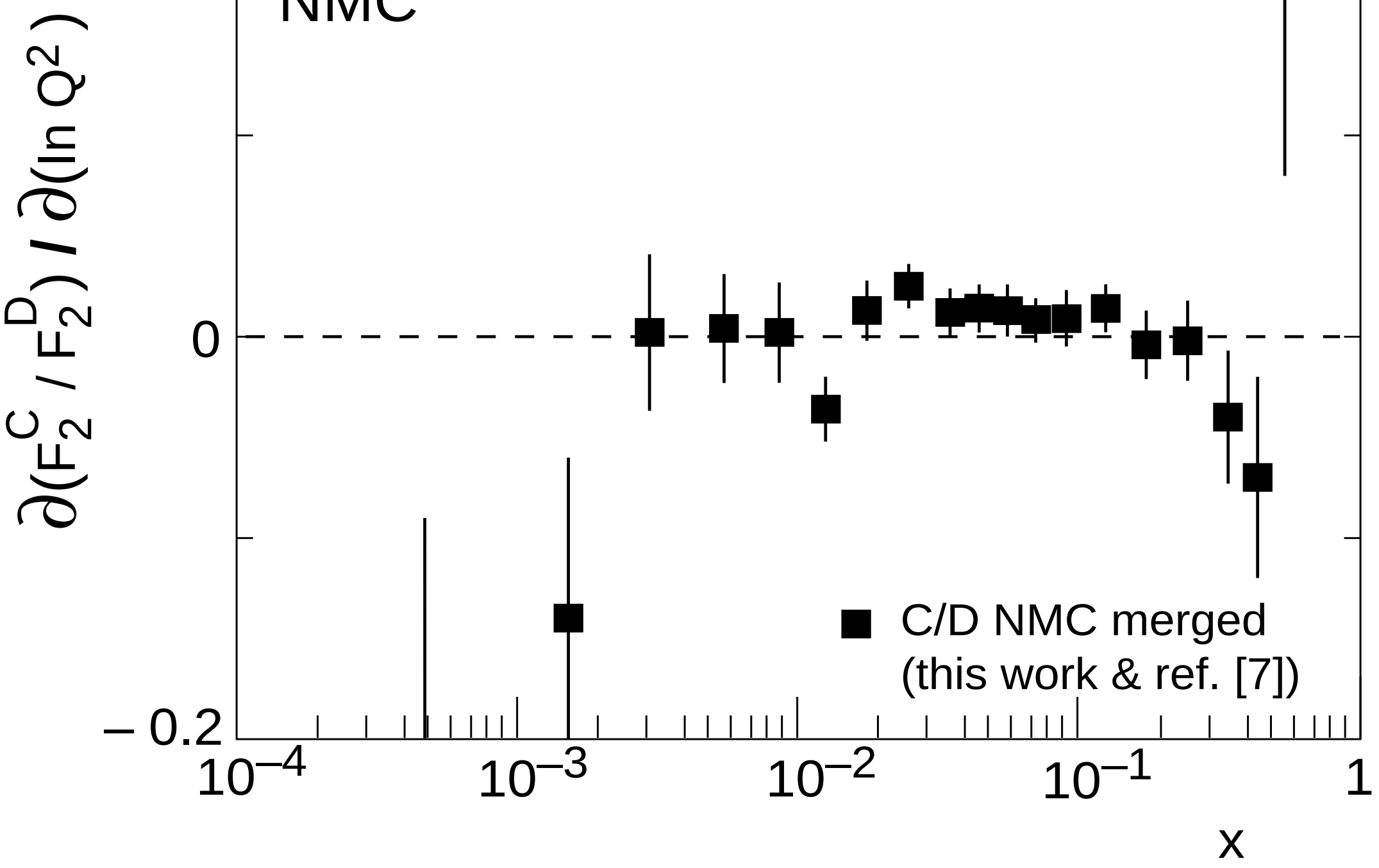

Figure 14: The logarithmic slopes $\partial(F_2^{\rm C}/F_2^{\rm D})/\partial(\ln Q^2)$ as obtained from linear least squares fits to the points of fig. 12.

## References

[1] CERN NA2/EMC, J.J. Aubert *et al.*, Phys. Lett. **123B** (1983) 275

[2] For a review of the experimental and theoretical situation see e.g. :
T. Sloan, G. Smadja and R. Voss, Phys. Rep. **162** (1988) 45;
L. Frankfurt and M. Strikman, Phys. Rep. **160** (1988) 235;
R.J.M. Covolan and E. Predazzi, in "Problems of Fundamental Modern Physics", Editors R. Cherubini, P. Dalpiaz and B. Minetti, World Scientific, Singapore (1991) p. 85;
M. Arneodo, Phys. Rep. **240** (1994) 301

[3] FNAL E665, M.R. Adams *et al.*, Phys. Rev. Lett. **68** (1992) 3266

[4] SLAC E139, R.G. Arnold *et al.*, Phys. Rev. Lett. **52** (1984) 727;
R.G. Arnold *et al.*, SLAC Report SLAC-PUB-3257 (1983);
J. Gomez *et al.*, Phys. Rev. D **49** (1994) 4348

[5] CERN NA37/NMC, P. Amaudruz *et al.*, Z. Phys. **C 51** (1991) 387

[6] CERN NA37/NMC, P. Amaudruz *et al.*, Z. Phys. **C 53** (1992) 73

[7] CERN NA37/NMC, M. Arneodo *et al.*, paper on reanalysis of [5] and [6] submitted to Nucl. Phys. B

[8] SLAC E140, S. Dasu *et al.*, Phys. Rev. Lett. **60** (1988) 2591

[10] A. Paic, Ph. D. thesis, Neuchatel University, 1994

[11] CERN NA37/NMC, P. Amaudruz *et al.*, Nucl. Phys. **B 371** (1992) 3

[12] CERN NA2/EMC, O.C. Allkhofer *et al.*, Nucl. Instr. Meth. **179** (1981)445-466

[13] A.A. Akhundov *et al.*, DESY 94-115;
A.A. Akhundov *et al.*, CERN-TH 7339/94;
A.A. Akhundov *et al.*, ICTP-IC/94/154

[14] CERN NA37/NMC, P. Amaudruz *et al.*, Phys. Letts **B295** (1992) 159

[15] B. Badełek and J. Kwieciński, Phys. Lett. **B 295** (1992) 263;
Z. Phys. **C 43** (1989) 251

[16] A. Donnachie, P. V. Landshoff, Phys. Lett. **B 296** (1992) 227;
A. Donnachie, P. V. Landshoff, Preprint M/C-TH 93/11

[17] M.D. Mestayer, Ph.D. thesis, Stanford University, SLAC-Report 214 (1978)

[18] L.W. Whitlow, SLAC-Report 357 (1990);
L.W. Whitlow *et al.*, Phys. Lett. **B 250** (1990) 193

[19] M. Gari and W. Krümpelmann, Z. Phys. **A 322** (1985) 689

[20] J. Bernabeu, Nucl. Phys. **B49** (1972) 186

[21] E.J. Moniz, Phys. Rev. **184** (1969) 1154

[22] A. Švarc and M.P. Locher, Fizika **22** (1990) 549;
M.P. Locher and A. Švarc, Z. Phys. **A 338** (1991) 89

[23] I. Sick, Phys. Lett. **116B** (1982) 212;
I. Sick, Nucl. Phys. **A218** (1974) 509;
I. Sick *et al.*, Phys. Lett. **88B** (1979) 245;
H. De Vries *et al.*, Atomic Data and Nuclear Data Tables, **36** (1987) 495

[24] C. Li et al., Nucl. Phys. **A 162** (1971) 583

[25] CERN NA2'/EMC, J. Ashman et al., Z. Phys. **C 52** (1991) 1

[26] W. T. Eadie, D. Drijard, F. E. James, M. Roos and B. Sadoulet, "Statistical Methods in Experimental Physics", North Holland 1971, p. 263

[27] ibid., pp. 282-3

[28] D.O. Caldwell *et al.*, Phys. Rev. Lett. **42** (1979) 553

[29] D. O. Caldwell *et al.*, Phys. Rev. **D 7** (1973) 1362